\documentclass[conference]{IEEEtran}
\IEEEoverridecommandlockouts
\usepackage{cite}
\usepackage{graphicx,color,epsfig,rotating}
\usepackage{amsfonts,amsmath,amssymb,bbm}
\usepackage{algorithm}
\usepackage{algpseudocode}
\usepackage{subfigure}
\usepackage{amsmath}
\usepackage{cite}
\usepackage{placeins}
\usepackage{graphicx}
\usepackage[latin1]{inputenc}
\usepackage{amssymb}
\usepackage{multirow}
\usepackage{stfloats}
\usepackage{tabularx} 
\usepackage{booktabs} 
\usepackage{url}
\usepackage{bm}
\usepackage{soul}
\usepackage{float}
\usepackage{lipsum}     
\usepackage{cuted} 
\usepackage{pstricks}
\usepackage{arydshln}
\usepackage{xspace} 

\allowdisplaybreaks  

\newcommand{\user}[1]{User~$#1$}

\long\def\comment#1{}

\makeatletter
\newcommand{\subalign}[1]{
  \vcenter{%
    \Let@ \restore@math@cr \default@tag
    \baselineskip\fontdimen10 \scriptfont\tw@
    \advance\baselineskip\fontdimen12 \scriptfont\tw@
    \lineskip\thr@@\fontdimen8 \scriptfont\thr@@
    \lineskiplimit\lineskip
    \ialign{\hfil$\m@th\scriptstyle##$&$\m@th\scriptstyle{}##$\crcr
      #1\crcr
    }%
  }
}
\makeatother

\newtheorem{example}{Example}
\newtheorem{theorem}{Theorem}
\newtheorem{lemma}{Lemma}

\DeclareMathOperator{\rank}{rank}  

\def \alphabm{{\bm{\alpha}}}
\def \alphabmstar{{\bm{\alpha}^*}}

\def \Fmbb{{\mbb{F}}}
\def \Fmbbq{{\mbb{F}_q}}
\def \Fmbbp{{\mbb{F}_p}}

\def \Wk{{W_k}}

\def \Zk{{Z_k}}

\def \rzsigma{{R_{Z_{\Sigma}}}}
\def \rzsigmastar{{R_{Z_{\Sigma}}^*}}
\def \rx{{R_X}} 

\def \rz{{R_Z}} 
\def \lz{{L_Z}}

\def \rxstar{{R_X^*}} 

\def \rzstar{{R_Z^*}}
\def  \zsigma{{Z_{\Sigma}}}
\def  \lzsigma{{L_{Z_{\Sigma}}}}

\def \dreg{{$d$-regular}\xspace}
\def \Amalpha{{\Am_\alphabm}}
\def \Amalphastar{{\Am_{\alphabm^*}}}

\let\trm\textrm
\let\tbf\textbf
\let\tit\textit

\let\mbf\mathbf
\let\mbb\mathbb

\let \bksl\backslash
\let \lf \left
\let  \rt \right

\newcommand{\kgm}{key generation matrix\xspace}
\newcommand{\compnts}{components\xspace}

\newcommand{\Secty}{Security\xspace}
\newcommand{\secty}{security\xspace}

\newcommand{\dimtns}{dimensions\xspace}
\newcommand{\dimtn}{dimension\xspace}

\newcommand{\suth}{such that\xspace}

\newcommand{\modltn}{modulation\xspace}

\newcommand{\ntwks}{networks\xspace}
\newcommand{\ntwk}{network\xspace}

\newcommand{\dmam}{diagonally modulated adjacency matrix\xspace}

\newcommand{\frmwk}{framework\xspace}
\newcommand{\Frmwk}{Framework\xspace}
\newcommand{\ntlztn}{neutralization\xspace}

\newcommand{\adjcy}{adjacency\xspace}

\newcommand{\cplt}{complete\xspace}
\newcommand{\Cplt}{Complete\xspace}

\newcommand{\nbrs}{neighbors\xspace}
\newcommand{\nbr}{neighbor\xspace}

\newcommand{\nbhd}{neighborhood\xspace}

\newcommand{\arbily}{arbitrarily\xspace}
\newcommand{\arbi}{arbitrary\xspace}

\newcommand{\Itf}{In the following\xspace}
\newcommand{\Thf}{Therefore\xspace}

\newcommand{\Aar}{As a result\xspace}

\newcommand{\brdcsts}{broadcasts\xspace}

\newcommand{\tsa}{topological secure aggregation\xspace}

\newcommand{\Decen}{Decentralized\xspace}
\newcommand{\decen}{decentralized\xspace}

\newcommand{\Inadd}{In addition\xspace}

\newcommand{\reqs}{requirements\xspace}
\newcommand{\req}{requirement\xspace}

\newcommand{\skr}{source key rate\xspace}

\newcommand{\topo}{topology\xspace}
\newcommand{\topos}{topologies\xspace}

\newcommand{\coeffts}{coefficients\xspace}
\newcommand{\coefft}{coefficient\xspace}

\newcommand{\ie}{i.e.\xspace}
\newcommand{\msg}{message\xspace}
\newcommand{\msgs}{messages\xspace}
\newcommand{\Wlog}{Without loss of generality\xspace}

\newcommand{\hie}{hierarchical\xspace}

\newcommand{\Msp}{More specifically\xspace}
\newcommand{\Ip}{In particular\xspace}

\newcommand{\af}{as follows\xspace}
\newcommand{\resp}{respectively\xspace}
\newcommand{\iid}{i.i.d.\xspace}
\newcommand{\thm}{theorem\xspace}
\newcommand{\Thm}{Theorem\xspace}
\newcommand{\schm}{scheme\xspace}

\newcommand{\info}{information\xspace}

\newcommand{\itic}{information-theoretic\xspace}

\newcommand{\etal}{\textit{et al.}\xspace}

\newcommand{\Bcuz}{Because\xspace}

\newcommand{\regu}{regular\xspace}

\newcommand{\agg}{aggregation\xspace}

\newcommand{\secagg}{secure aggregation\xspace}

\newcommand{\diffce}{difference\xspace}
\newcommand{\Diff}{Different\xspace}

\newcommand{\Fex}{For example\xspace}
\newcommand{\indep}{independent\xspace}

\newcommand{\indepce}{independence\xspace}

\newcommand{\indiv}{individual\xspace}

\newcommand{\Indiv}{Individual\xspace}

\newcommand{\comm}{communication\xspace}

\newcommand{\Comm}{Communication\xspace}
\newcommand{\achv}{achieve\xspace}

\newcommand{\achvb}{achievable\xspace}
\newcommand{\Achvb}{Achievable\xspace}
\newcommand{\achvblty}{achievability\xspace}

\newcommand{\distn}{distribution\xspace}

\newcommand{\muinfo}{mutual information\xspace}

\newfont{\bbb}{msbm10 scaled 700}

\newfont{\bb}{msbm10 scaled 1100}

\newcommand{\bv}{{\mathbf{b}}}

\newcommand{\vv}{{\mathbf{v}}}

\newcommand{\kth}{{$k^{\rm th}$ }}

\newcommand{\Am}{{\bf A}}

\newcommand{\Hm}{{\bf H}}

\newcommand{\Ac}{{\cal A}}
\newcommand{\Bc}{{\cal B}}

\newcommand{\Ec}{{\cal E}}

\newcommand{\Gc}{{\cal G}}

\newcommand{\Ic}{{\cal I}}

\newcommand{\Nc}{{\cal N}}

\newcommand{\Rc}{{\cal R}}

\newcommand{\Vc}{{\cal V}}

\newcommand{\transfm}{transform\xspace}
\newcommand{\inputsumnbrk}{\hbox{$\sum_{i\in \Nc_k}W_i$}}
\newcommand{\inputsetnbrk}{\hbox{$\{W_i\}_{i\in \Nc_k}$}}

\newcommand{\msgsumnbrk}{\hbox{$\sum_{i\in \Nc_k}X_i$}}
\newcommand{\msgsetnbrk}{\hbox{$\{X_i\}_{i\in \Nc_k}$}}

\newcommand{\keysumnbrk}{\hbox{$\sum_{i\in \Nc_k}Z_i$}}

\newcommand{\diag}{{\mathrm{diag}}  }

\newcommand{\eqdef}{\stackrel{\Delta}{=}}

\newcommand{\be}{\begin{equation}}
\newcommand{\ee}{\end{equation}}
\newcommand{\bea}{\begin{eqnarray}}
\newcommand{\eea}{\end{eqnarray}}

\newcommand{\besub}{\begin{subequations}}
\newcommand{\eesub}{\end{subequations}}

\begin{document}
\title{Information-Theoretic Secure Aggregation over Regular Graphs}

\author{\IEEEauthorblockN{Xiang Zhang\IEEEauthorrefmark{1}, 
Zhou Li\IEEEauthorrefmark{2},
Han Yu\IEEEauthorrefmark{1}, 
Kai Wan\IEEEauthorrefmark{3}, 
Hua Sun\IEEEauthorrefmark{4}, 
Mingyue Ji\IEEEauthorrefmark{5}, 
and Giuseppe Caire\IEEEauthorrefmark{1}
}
\IEEEauthorblockA{
Department of Electrical Engineering and Computer Science, Technical University of Berlin\IEEEauthorrefmark{1}\\
Guangxi Key Laboratory of Multimedia Communications and Network Technology, Guangxi University\IEEEauthorrefmark{2}\\
School of Electronic Information and Communications, Huazhong University of Science and Technology\IEEEauthorrefmark{3}\\
Department of Electrical Engineering, University of North Texas\IEEEauthorrefmark{4}\\
Department of Electrical and Computer Engineering, University of Florida\IEEEauthorrefmark{5}\\
Email:~\IEEEauthorrefmark{1}\{xiang.zhang, caire\}@tu-berlin.de,~\IEEEauthorrefmark{1}han.yu.1@campus.tu-berlin.de,~\IEEEauthorrefmark{2}lizhou@gxu.edu.cn,}
\IEEEauthorblockA{~\IEEEauthorrefmark{3}kai\_wan@hust.edu.cn,~\IEEEauthorrefmark{4}hua.sun@unt.edu,~\IEEEauthorrefmark{5}mingyueji@ufl.edu}
}

\maketitle

\begin{abstract} 
Large-scale decentralized learning frameworks such as federated learning (FL), require both communication efficiency and strong data security, motivating the study of secure aggregation (SA). While information-theoretic SA is well understood in centralized and fully connected networks, its extension to decentralized networks with limited local connectivity remains largely unexplored.
This paper introduces \emph{topological secure aggregation} (TSA), which studies one-shot, information-theoretically secure aggregation of neighboring users' inputs over arbitrary network topologies. 
We develop a unified linear design framework that characterizes TSA achievability through the spectral properties of the communication graph, specifically the kernel of a diagonally modulated adjacency matrix. 
For several representative classes of 
$d$-regular graphs including ring, prism and complete topologies, we establish the optimal communication and secret key rate region. In particular, to securely compute one symbol of the neighborhood sum, each user must (i) store at least one key symbol, (ii) broadcast at least one message symbol, and (iii) collectively, all users must hold at least 
$d$ i.i.d. key symbols. Notably, this total key requirement depends only on the \emph{neighborhood size} 
$d$, independent of the network size, revealing a fundamental limit of SA in decentralized networks with limited local connectivity.
\end{abstract}

\section{Introduction}
\label{sec:intro} 
Federated learning (FL)  enables collaborative training of  machine learning (ML) models without directly  exchanging the clients' private datasets~\cite{mcmahan2017communication}.
In the centralized architecture,
a parameter server coordinates the model \agg  process, whereas in decentralized FL (DFL)~\cite{lalitha2018fully, beltran2023decentralized}, model aggregation is distributed across neighboring nodes over a communication graph.
Even though the local datasets are not explicitly shared, FL still exposes vulnerabilities in the presence of an untrusted server or peers~\cite{bouacida2021vulnerabilities, geiping2020inverting, mothukuri2021survey}.
The demand for stronger security guarantees has motivated the research of \secagg (SA) in FL~\cite{bonawitz2017practical, bonawitz2016practical,wei2020federated,hu2020personalized,zhao2020local,so2021turbo}. 
\Fex, Bonawitz \etal~\cite{bonawitz2017practical,bonawitz2016practical}
proposed an SA protocol relying on pairwise secret key masking to protect the users' local models from an untrusted server. Since then, a large body of SA algorithms based on cryptographic primitives  has been proposed, typically focusing on computational security guarantees. 

In contrast, \itic \secagg~\cite{9834981, zhao2023secure,so2022lightsecagg,jahani2022swiftagg,jahani2023swiftagg+,wan2024information,wan2024capacity, zhang2025secure,zhao2022mds, zhao2023optimal,li2023weakly,li2025weakly,li2025collusionresilienthierarchicalsecureaggregation,sun2023secure, yuan2025vector,zhang2024optimal, 10806947,egger2024privateaggregationhierarchicalwireless, lu2024capacity,zhang2025fundamental, xu2025hierarchical, 11195652,li2025optimal, li2025capacity, li2026optimal,zhang2025information}
studies the fundamental limits of SA, including both communication efficiency and secret key generation efficiency, with information-theoretic \tit{perfect} security characterized by a zero mutual information criterion. The original formulation of Zhao and Sun\cite{9834981, zhao2023secure} studies SA in a server-based star network with $K$ users, whose local FL models are represented by a set of \tit{input} variables $W_1,\cdots, W_K$. The server aims to recover  the sum  of  the inputs $W_1+ \cdots+W_K$ of the users based on their uploaded \msgs $X_1, \cdots, X_K$. \Secty, as represented by
$ I(\{X_k\}_{k=1}^K; \{W_k\}_{k=1}^K|\sum_{k=1}^KW_k)=0
$, 
requires that the untrusted server infer no \info about the  inputs  of the users beyond their aggregate sum. 
Following this formulation, SA has been studied under a multitude of constraints, including user dropout and collusion resilience~\cite{9834981, zhao2023secure,zhang2025secure,so2022lightsecagg,jahani2022swiftagg,jahani2023swiftagg+}, groupwise keys~\cite{zhao2023secure,wan2024information,wan2024capacity, zhang2025secure,li2025capacity}, user selection~\cite{zhao2022mds, zhao2023optimal}, heterogeneous security models~\cite{li2023weakly,li2025weakly,li2025collusionresilienthierarchicalsecureaggregation,li2025optimal}, oblivious server~\cite{sun2023secure},  multiple recovery objectives~\cite{yuan2025vector},  \hie SA (HSA)~\cite{zhang2024optimal, 10806947,egger2024privateaggregationhierarchicalwireless, lu2024capacity,zhang2025fundamental, 11195652,li2025collusionresilienthierarchicalsecureaggregation,xu2025hierarchical,weng2025cooperative}, \decen SA (DSA)\cite{zhang2025information,li2025capacity,li2025optimal}, and multi-server SA\cite{li2026optimal}.

In this work, we study \tit{\tsa} (TSA), which formulates the one-shot, \info-theoretically secure \agg  of  neighboring users' inputs in \decen networks with \arbi  \topos. TSA can be viewed as an extension of \decen SA (DSA)\cite{zhang2025information}, which studies SA in fully connected networks, to more general \topos with uneven and limited \nbhd connectivity. 
We propose a unified linear design \frmwk that captures the simultaneous input sum recovery and \secty requirements at all users through a kernel-based representation of a \dmam of the underlying \comm graph.
Under this design \frmwk, we identify several classes of $d$-\regu graphs, including the ring, the prism graph, and the \cplt graph, that achieve the optimal rate region $\Rc^*=\{(\rx, \rz, \rzsigma): \rx\ge 1,  \rz  \ge 1, \rzsigma \ge d  \}$ where $\rx, \rz$ and $\rzsigma$ denote the \comm rate, \indiv secret key rate, and the total (source) key rate, \resp. 
The characterization of $\Rc^*$ consists of  a set of novel secret key and broadcast \msg designs, together with a nontrivial converse proof that establishes tight lower bounds on the communication and key rates.

\emph{Notation.}
$[n] \eqdef \{1,  \cdots, n\} $,
$\Ac\bksl \Bc  \eqdef \{x\in  \Ac:  x \notin  \Bc \}$. $\Ac\bksl \Bc$ is written as $\Ac\bksl  b$ if $\Bc=\{b\}$. For an index  set $\Ac$, let $X_\Ac \eqdef \{X_i: i\in \Ac\} $. $X_\Ac$ is written as $X_{1:n}$ if $\Ac=\{1,\cdots,n\}$. $m \mid n$ means $m$ divides $n$.
$\underline{x}_{m\times n}$ denotes the $m\times n$ matrix with all entries equal  to $x$.
For matrix $\Am$, $\Am_{\Ic,:}$ denotes the submatrix of $\Am$ consisting of the rows indexed by $\Ic$.

\section{Problem Formulation}
\label{sec:problem formulation}
Consider  a  \comm  network represented  by an  undirected, connected graph $\Gc=(\Vc, \Ec)$, where the vertex set $\Vc $ represents $K$ users and the edge set $\Ec$ represents the \comm links between the users. 
Let $\Nc_k \subseteq [K]\bksl \{k\} $  denote the neighbor set of user $k$. 
User $k$ holds an \emph{input} variable $W_k \in \mbb{F}_q^{1\times L}$ which
consists  of $L$ \iid uniform symbols from some finite field $\mbb{F}_q$, representing locally trained models/gradient updates in federated learning. 
Different users are assumed to be \indep\footnote{It should be noted that the proposed \secagg scheme works for \arbily distributed and correlated inputs. The
\iid assumption on the inputs is only necessary to establish the optimality of the scheme. 
}, \ie,
\begin{align}
\label{eq:input indep}
H(W_k) & = L \quad (\trm{in $q$-ary  units}),\notag\\
H(W_{1:K}) & = \sum\nolimits_{k=1}^K H(W_k).
\end{align}
To achieve \secagg, each user is also assigned a secret key $Z_k$ consisting of $L_Z$ symbols from  $\mbb{F}_q$.
To quantify the amount of randomness consumed   by the keys, a \emph{source key} variable $\zsigma$---consisting of  $\lzsigma$ symbols---is introduced, from which the \indiv keys $Z_{1:K}$ are derived, \ie, 
\be
\label{eq:key gen from source key}
H\left(Z_{1:K} | \zsigma \right)=0.
\ee 
Similar to existing literature~\cite{9834981,zhao2023secure, zhao2023optimal,zhang2024optimal,zhang2025fundamental},  the secret keys are assumed to be generated and distributed privately to each user by a trusted third-party entity. 
Moreover,  the keys are \indep of the inputs, 
\be
\label{eq:indep between inputs and keys}
H\left(W_{1:K},Z_{1:K}\right) = \sum\nolimits_{k=1}^K H(W_k)+ H\left(Z_{1:K}\right).
\ee

\tbf{\Comm protocol.}
Each user is connected to its neighbors through an error-free broadcast channel.
The broadcast transmissions of different users are assumed to be orthogonal.
In the considered \secagg problem, each user aims to recover the sum of the inputs within its neighborhood. To this end,  User $k$ generates a \msg $X_k$ containing $L_X$ symbols, which is broadcast to all users in $\Nc_k$. $X_k$ is
a deterministic function of the input $W_k$ and key $Z_k$, such that
\be
\label{eq:msg gen}
H(X_k|W_k, Z_k)=0.
\ee 
 
\tbf{Goal.}
After exchanging the \msgs, each user should be able to  recover the  sum of the inputs from all of its neighbors (including itself), \ie, 
\begin{align}
\label{eq:recovery} 
H\left(\sum\nolimits_{i\in \Nc_k \cup \{k\}  }W_i \Big| X_{\Nc_k}, W_k, Z_k  \right)=0, \;\forall k\in[K]
\end{align}
Note that recovering $\inputsumnbrk$  suffices to ensure the above recovery constraint since $W_k$ is already  available to User $k$.

\tbf{Security model.}
We consider the \emph{honest-but-curious} security model commonly used in secure FL and  multi-party computation (MPC)~\cite{bonawitz2017practical, goldreich1998secure,jahani2023swiftagg+}, where each user executes the aggregation protocol faithfully (\ie, does not tamper with its inputs) but may attempt to infer its neighbors'  private inputs from the observed \msgs.
\Ip, security requires that each user must not infer any information about the inputs of other users---essentially those of its neighbors, since it cannot observe messages outside its neighborhood---beyond the aggregate sum of inputs $\inputsumnbrk  $ within its neighborhood $\Nc_k$. 
\Msp, security is satisfied if the observed \msgs $\{X_i\}_{i\in \Nc_k}$  are statistically independent of the inputs  $\{W_i\}_{i\in \Nc_k}$ at each \nbhd,
\ie, 
\begin{align}
\label{eq:security}
& I\left(X_{\Nc_k};W_{\Nc_k} \Big| \sum\nolimits_{i\in \Nc_k}W_i , W_k, Z_k \right)=0, \; \forall k\in[K]
\end{align}
The  conditioning terms in (\ref{eq:security}) arise from the fact that User $k$ must eventually recover $\sum_{i\in \Nc_k}W_i$, while   $W_k$ and $Z_k$ are naturally available to  User $k$.

A \secagg scheme consists of a design of the secret keys $\zsigma,Z_1, \cdots, Z_K$ (see (\ref{eq:key gen from source key})) and the broadcast \msgs $X_1,\cdots,X_K$ (see (\ref{eq:msg gen})) such that all users can correctly recover the desired input sum (\ref{eq:recovery}) while adhering to the security constraint (\ref{eq:security}).
Both \comm efficiency and secret key generation  efficiency, measured by the \comm and key rates, \resp, are considered.  Specifically, the \comm rate $\rx$, \indiv key rate $\rz$, and source  key rate $\rzsigma$ are \resp defined as
\be
\label{eq:def Rx,RZ,RZSigma}
\rx \eqdef   {L_X}/{L},  \; 
\rz \eqdef    {L_Z}/{L},  \; 
\rzsigma \eqdef    {\lzsigma}/{L}. 
\ee 
A rate tuple $(\rx, \rz,\rzsigma)$ is said to be achievable if there exists a  scheme that simultaneously  achieves  the rates $\rx,\rz$ and $\rzsigma$. We aim to characterize the optimal rate region  $\Rc^*$, defined as the closure of all achievable rate tuples.

\section{Main Result}
\label{sec:main result}

\begin{theorem}
\label{thm:main result}
\tit{For \secagg over  $d$-\regu networks, including the ring network ($d=2$), the prism graph ($d=3$), and the complete graph ($d=K-1$), 
the optimal rate region is given by}
\be
\label{eq:optiaml region,main thm}
\Rc^*=\lf\{(\rx, \rz, \rzsigma): \rx \ge  1, \rz \ge 1 , \rzsigma \ge d \rt \}.
\ee 
\end{theorem}

The \achvblty and converse proofs of \Thm \ref{thm:main result} are given in Sections \ref{sec:proposed scheme} and \ref{sec:converse}, \resp. For \achvblty, we present a linear design \frmwk in which the keys are constructed from the basis of the kernel space of a diagonally modulated \adjcy matrix (DMAM) of the  \comm graph $\Gc$. 
We also propose a novel entropic converse proof that establishes the optimality of the proposed scheme. 

We highlight the implications  of  the  \thm  \af:

1) \emph{Source Key Efficiency}:
Notably, the optimal \skr  $\rzsigmastar=d$ depends solely on the   \nbhd size  $d$, rather than the total number of users $K$. This result  establishes a fundamental limit for \decen \secagg in networks with limited local connectivity. Although \Thm \ref{thm:main result} enumerates several classes of $d$-\regu  graphs for which $\rzsigmastar=d$ is \achvb, there might exist other $d$-\regu graphs that cannot \achv the extremal rate region (\ref{eq:optiaml region,main thm}). Under the proposed one-shot linear design \frmwk in Section \ref{sec:linear design frmwk}, \achvblty of (\ref{eq:optiaml region,main thm}) depends on whether the DMAM of $\Gc$---which captures the key-cancellation conditions required for input sum recovery---admits a kernel of dimension at least $d$.
Since \secty requires at least $d$ \indep source key dimensions, if the DMAM provides fewer than $d$ kernel \dimtns, the secret keys cannot be fully neutralized, thereby preventing correct recovery of the input sums.

2) \emph{Connection to \Decen SA (DSA)~\cite{zhang2025information}}:
SA has been previously studied in a fully connected \ntwk, referred to as \decen SA (DSA)~\cite{zhang2025information}. DSA can be viewed as a special instance of TSA in which the underlying \comm network is a complete graph ($(K-1)$-\regu). Indeed, the optimal rate region is given by $\Rc_{\rm DSA}^*=\{(\rx,\rz,\rzsigma):\rx \ge 1, \rx\ge 1, \rzsigma \ge  K-1\}$, which follows directly from \Thm \ref{thm:main result}. Refer to Section \ref{subsec:H specification} for a detailed explanation.

\section{Motivating Examples}
\label{sec:examples}

\begin{example}[Prism  Graph]
\label{example:K=6 prism graph}
Consider the $6$-user prism graph as shown in Fig.~\ref{fig:prism graph,example}. 
\begin{figure}[t]
    \centering
    \includegraphics[width=0.15\textwidth]{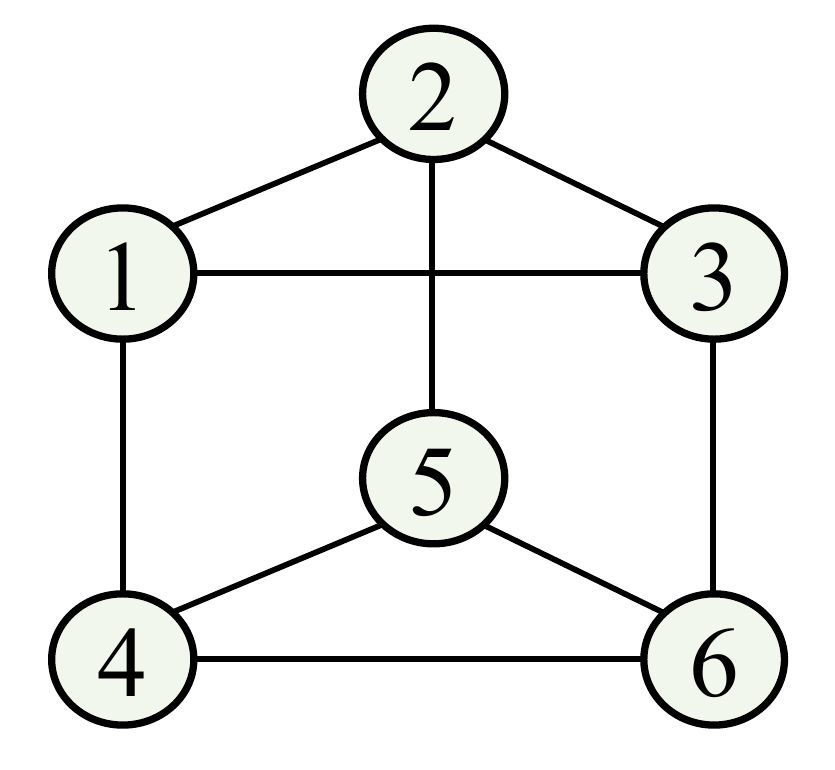}
    \vspace{-0.4cm}
    \caption{Prism graph with $6$  users each having  $3$  neighbors. \user{1} wants to recover $W_1+W_2+W_3+W_4$.}
    \label{fig:prism graph,example}
    \vspace{-0.5cm}
\end{figure}
Suppose each input contains $L=1$ symbol from  $\mbb{F}_5$. The design  of the secret keys and broadcast \msgs are described \af.

\tbf{Secret key design.}
Let the source key be
\be
\label{eq:souce key, example}
\zsigma =(N_1,N_2, N_3)
\ee
where $N_1, N_2,N_3$ are $3$ \iid uniform random symbols from $ \mbb{F}_5$, implying a source key rate of $\rzsigma=3$. 
The \indiv secret keys are chosen as
\begin{eqnarray}
\label{eq:indiv keys,prism example}
Z_1  =N_1, &  Z_4  = - ( 2 N_1 + N_2+N_3 ), \notag\\
Z_2    =N_2, & Z_5   =-(N_1 + 2 N_2 +N_3 ),\notag\\
Z_3    =N_3, & Z_6     = -(N_1+N_2+ 2N_3).
\end{eqnarray}
Since  each key is a  linear combination of the source key symbols, the \indiv  key rate is equal to $\rz=1$. 

\tbf{Message design.}
Each broadcast \msg is simply the sum of the  input and the corresponding key, \ie, 
\be 
\label{eq:messages,prism example}
X_k= W_k + Z_k, \;k \in[6]
\ee 
User $k$ then broadcasts $X_k$ to all its \nbrs in $\Nc_k$, implying  $\rx=1$ because each \msg contains one symbol.

\tbf{Proof of recovery.} 
\user{k} computes the sum of  $2Z_k$ and all received \msgs $\{X_i\}_{i\in \Nc_k}$  to recover the desired input sum \inputsumnbrk:
$2Z_k + \msgsumnbrk   \overset{(\ref{eq:messages,prism example})}{=}
\inputsumnbrk + \keysumnbrk + 2Z_k 
 \overset{(\ref{eq:indiv keys,prism example})}{=} \inputsumnbrk$, 
where the second equality is due to the zero-sum property of the secret keys in (\ref{eq:indiv keys,prism example}),
\be
2Z_k + \keysumnbrk =0,\; \forall k \in[6]
\ee 
\Fex, for \user{1} with \nbr set $\Nc_1=\{2,3,4\}$,  we have $2Z_1 +Z_2 +Z_3+Z_4=2N_1+N_2 +N_3 -(2N_1+N_2+N_3)=0$. For \user{4} with $\Nc_4=\{1,5,6\}$, we have  $2Z_4 +Z_1 +Z_5+Z_6=-5(N_1+N_2+N_3)=0 $. Importantly, the zero-sum property of the secret design
(\ref{eq:indiv keys,prism example}) ensures that each user's own key $Z_k$ and the key \compnts contained in the received \msgs perfectly cancel out when added,  yielding the desired input sum.

\tbf{Proof of security.}
\Secty (\ref{eq:security}) requires that, given  what is known (\ie, $\Wk, \Zk$) or is recoverable  (\ie, \inputsumnbrk) by \user{k}, the set of \nbhd inputs \inputsetnbrk should be \indep of the  observed \msgs $\msgsetnbrk$. \Itf, we show that \secty is indeed satisfied at all users by first providing an intuitive proof based on linear decoding, followed by a rigorous proof using \muinfo.

\tit{Intuitive proof}: Suppose \user{k} applies a linear \transfm $\bm{l}_k\eqdef (a_k,b_k,c_k,d_k)\in \Fmbb_5^{1\times 4}$ to the variables $(Z_k  ,\msgsetnbrk)$ to recover $\inputsumnbrk$. \Wlog, let us consider \user{1} which applies $\bm{l}=(a,b,c,d)$ to $(Z_1,X_2,X_3,X_4)$,
\begin{align}
&(Z_1,X_2,X_3,X_4) \bm{l}^T= aN_1+ b(W_2+N_2) + c(W_3+N_3)\notag\\
& \hspace{3.4cm}  + d(W_4-(2N_1+N_2+N_3))\notag\\
& = bW_2+ cW_3+dW_4 +(a-2d)N_1+(b-d)N_2\notag\\
& \hspace{5.4cm}+(c-d)N_3.
\end{align}
To neutralize the unknown key symbols $N_2$ and $N_3$ (as well as the known symbol $N_1$), it must hold that $a=2d,b=c=d$. Consequently,  the linear \transfm must take the form $\bm{l}=(2a, a,a,a)$, yielding  $(Z_1,X_2,X_3,X_4) \bm{l}^T=a(W_2+W_3+W_4)$, which is a {function} of the desired input sum $W_2+W_3+W_4$. 
This shows that,  the inputs remain protected if the unknown key  symbols $N_2,N_3$ are not neutralized. Otherwise, if $N_2$ and $N_3$ perfectly cancel out,  the only \info about $(W_2,W_3,W_4)$  revealed to \user{1} is the desired aggregate $W_2+W_3+W_4$, which is intended to be  recoverable  by \user{1}.
In either case, nothing beyond the desired input sum is leaked to \user{1}, thereby ensuring \secty at \user{1}.

\tit{Formal proof}:
Beyond the above intuitive proof, a rigorous proof of \secty is provided \af.  Still consider \user{1}. Let us denote  $W_{2+3+4}\eqdef W_2+W_3+W_4$.  
We have
\begin{subequations}
\label{eq:formal security proof, example}
\begin{align}
& I\lf(X_{\{2,3,4\}} ; W_{\{2,3,4\}}|W_1,Z_1, W_{2+3+4}\rt)\notag\\
& \overset{(\ref{eq:input indep}),(\ref{eq:indep between inputs and keys}),(\ref{eq:msg gen})  }{=}   I\lf(X_{\{2,3,4\}} ; W_{\{2,3,4\}}|Z_1, W_{2+3+4}\rt) 
\label{eq:step0}\\
&\quad \;\,   =H\lf(X_{\{2,3,4\}}|Z_1, W_{2+3+4} \rt)\notag\\
& \qquad \;\;  -H\lf(X_{\{2,3,4\}}|Z_1, W_{2+3+4},W_{\{2,3,4\}} \rt),
\label{eq:step1}
\end{align}
\end{subequations}
where (\ref{eq:step0}) is because $W_1$ is \indep of all the remaining terms. 
The first term  in  (\ref{eq:step1}) can be upper bounded by
\begin{subequations}
\label{eq:first term, example}
\begin{align}
&  H\lf(X_{\{2,3,4\}}|Z_1, W_{2+3+4} \rt ) \notag \\
&  = H\big( \{W_i+Z_i\}_{i=2}^4, W_{2+3+4} |Z_1\big)  - H(W_{2+3+4}|Z_1)\\
&  \overset{(\ref{eq:indiv keys,prism example}),(\ref{eq:indep between inputs and keys})}{=}
H\big(\{W_i+N_i\}_{i=2}^3, W_4-(2N_1+N_2+N_3),\notag\\
& \hspace{3.5cm}   W_{2+3+4}  |N_1\big)  - H(W_{2+3+4})
\label{step0, 1st term, example}
\\
& \overset{(\ref{eq:input indep})}{=}
H\big(\{W_i+N_i\}_{i=2}^3, W_4-(N_2+N_3) |N_1\big)  - 1
\label{step1, 1st term, example}
\\
& \overset{(\ref{eq:indep between inputs and keys}),(\ref{eq:souce key, example})}{=}
H(W_2+N_2,W_3+N_3, W_4-(N_2+N_3))  - 1
\label{step2, 1st term, example}
\\
& \; \le 3-1=2, \label{step3, 1st term, example}
\end{align}
\end{subequations}
where (\ref{step0, 1st term, example}) follows from the \indepce between  the inputs and keys (see (\ref{eq:indep between inputs and keys})). (\ref{step1, 1st term, example}) holds because $W_2+W_3+W_4$  can be recovered from $W_2+N_2,W_3+N_3$ and $W_4-(N_2+N_3)$, whose sum equals $W_2+W_3+W_4$. (\ref{step2, 1st term, example}) is due to the \indepce between the keys and inputs, and the \indepce among the source  key symbols. 
Finally, (\ref{step3, 1st term, example}) holds because each \msg contains one symbol, and an \iid uniform \distn maximizes the entropy.
The  second term in (\ref{eq:step1}) is equal to 
\begin{subequations}
\label{eq:second term, example}
\begin{align}
& H\big(X_{\{2,3,4\} }|Z_1, W_{2+3+4},W_{2,3,4} \big )\notag\\
& \overset{ (\ref{eq:messages,prism example})  }{=} 
H\big(\{W_i+Z_i\}_{i=2}^4|Z_1, W_{2,3,4} \big)
\label{step0, second term, example}
\\
& = 
H\big(Z_2, Z_3,Z_4|Z_1, W_{2,3,4} \big)\\
& \overset{(\ref{eq:indiv keys,prism example})  }{=}
H(N_2, N_3, -(2N_1+N_2+N_3)  |N_1, W_{2,3,4} )
\label{step1, second term, example}
\\
& \overset{(\ref{eq:indep between inputs and keys}),(\ref{eq:souce key, example})}{=}
H(N_2, N_3, -(N_2+N_3))=H(N_2, N_3)=2,\label{step3, second term, example}
\end{align}
\end{subequations}
where we plugged in the \msg and  key designs in (\ref{step0, second term, example}) and (\ref{step1, second term, example}). (\ref{step3, second term, example}) is due  to the \indepce and uniformity of the source key symbols.
Plugging (\ref{eq:first term, example}) and  (\ref{eq:second term, example})  into  (\ref{eq:formal security proof, example}),  and because mutual 
\info  is nonnegative, we have $I(X_{\{2,3,4\}  } ; W_{\{2,3,4\}  }|W_1,Z_1, W_{2+3+4}   )=0$, proving  security at \user{1}. Security can be proved similarly for other users.
\hfill $\lozenge$
\end{example}

\section{\Achvb Scheme}
\label{sec:proposed scheme}

In this section, we present a unified linear design framework that achieves the optimal rate region  $\Rc^*$ (\ref{eq:optiaml region,main thm}) established in \Thm  \ref{thm:main result}. 
The proposed \frmwk consists of two main components: 1) a kernel-based representation of  the secret key \ntlztn and input sum recovery through the \dmam of the  \comm graph $\Gc$, and 2) a set of  \nbhd rank conditions on the key generation matrix that ensure \secty. These components are described in the following subsections.

\subsection{The One-Shot Linear Design \Frmwk}
\label{sec:linear design frmwk}
Consider any \dreg graph $\Gc$ and the optimal rates of $\rxstar=1, \rzstar=1$ and $\rzsigmastar=d$. Suppose each input has $L=1$ symbol from a proper field $\Fmbb_q$. Thus, each  \indiv key has $\lz=1$ symbol, $ Z_k \in \Fmbb_q$, and the source key $\zsigma=(N_1, \cdots, N_d)$ contains $\lzsigma=d$ \iid symbols $N_1, \cdots, N_d$.

\tbf{Secret key design.} 
The \indiv keys are generated as
\be
\label{eq:indiv key design, one-shot framework}
(Z_1, \cdots, Z_K)^T = \Hm(N_1, \cdots, N_d)^T, 
\ee
where $\Hm \in \Fmbb_q^{K \times d}$ denotes the key generation matrix. Each \indiv key $Z_k=\Hm_{k,:}Z_\Sigma^T  $ is a linear combination of the source key  symbols $N_1, \cdots, N_d$ and contains $\lz=1$ symbol.

\tbf{Message design.} 
User $k$ broadcasts one symbol
\be
\label{eq:message design, one-shot framework}
X_k= W_k + Z_k
\ee
to its \nbrs $\Nc_k,\forall k\in [K] $, yielding a \comm rate of  $\rx=1$. Accordingly, each \user{k} receives a total of $d$ \msgs $\{X_i\}_{i\in \Nc_k}$  from its \nbrs.

\tbf{Input sum recovery.} \user{k}   recovers the desired sum of \nbhd inputs $\sum_{i\in  \Nc_k}W_i$ by linearly combining its own secret key $Z_k$ (with \coefft $\alpha_k$) and the received \msgs : 
\begin{align}
\label{eq:linear recovery, one-shot framework}
& \sum\nolimits_{i\in \Nc_k}W_i   = \alpha_k Z_k + \sum\nolimits_{i\in \Nc_k}X_i \notag    \\
 &  \hspace{1.4cm} \overset{(\ref{eq:message design, one-shot framework})}{=}  \sum\nolimits_{i\in \Nc_k}W_i + \lf(\alpha_k Z_k + \sum\nolimits_{i\in \Nc_k}Z_i\rt).
\end{align}
For (\ref{eq:linear recovery, one-shot framework}) to hold,  the secret key components
$Z_k,\{Z_i\}_{i\in {\Nc_k}}$
must cancel out simultaneously  at all users, \ie,
\be
\label{eq:key neutralization, one-shot framework}
\alpha_k Z_k + \sum\nolimits_{i\in \Nc_k}Z_i=0, \; \forall k\in  [K]
\ee 
Let $\Am \in  \{0,1\}^{K\times K}$ be the \adjcy matrix of $\Gc$, and let $\alphabm\eqdef (\alpha_1, \cdots, \alpha_K)$ 
denote the \tit{key \ntlztn \coeffts} used by all users. Then (\ref{eq:key neutralization, one-shot framework}) can be written in matrix form as
\be
\label{eq:key neutralization, matrix form, one-shot framework}
\left(\diag(\alphabm)+ \Am\right)\Hm(N_1, \cdots, N_d)^T=\mbf{0}_{K\times 1}.
\ee 
Since the source key symbols are \iid uniform, (\ref{eq:key neutralization, matrix form, one-shot framework}) holds iff
\be
\label{eq:key neutralization, no Ns, one-shot framework}
\left(\diag(\alphabm)+ \Am\right)\Hm= \mbf{0}_{K\times  d}.
\ee 
For ease of notation, we define $\Am_\alphabm  \eqdef \diag(\alphabm)+ \Am$
and  refer to $\Amalpha$ as  the \tit{\dmam} (DMAM) of $\Gc$, where the diagonal perturbation of $\alphabm$ on the \adjcy matrix $\Am$ is termed a \emph{modulation}.
Equation (\ref{eq:key neutralization, no Ns, one-shot framework})  then becomes
\be
\label{eq: Aalpha*H=0} 
\mathrm{[Recovery]}\quad 
\Amalpha \Hm= \mbf{0}_{K\times d}.
\ee 
Essentially,  (\ref{eq: Aalpha*H=0}) requires  $\Hm$ be designed such that there exists a 
key \ntlztn \coefft vector $\alphabm$  for which $\Hm $ lies in the kernel space of $\Amalpha$, \ie, $\Hm \subseteq \ker(\Amalpha)  $.

\tbf{Security.} 
A necessary condition for achieving security is that {each set of $d+1$ \nbhd keys must contain at least $d$ \indep key symbols as shown in the following lemma.
\begin{lemma}
\label{lemma:nbhd d+1 indep symbols, necessary cond for sec, oneshot frmwk}
\tit{For the \secty constraint in (\ref{eq:security}) to hold, the secret keys must satisfy (with all entropies in $q$-ary units)}
\besub
\label{eq:neighborhood key d/d-1 symbols}
\begin{align}
H\left( Z_{\Nc_k  \cup  \{k\} }   \right)  & \ge d, 
\label{eq:d, neighborhood key d/d-1 symbols}
\\
H\left( Z_{\Nc_k } |Z_k   \right)  & \ge d-1,\;
\forall k\in [K]
\label{eq:d-1, neighborhood key d/d-1 symbols}
\end{align}
\eesub
\end{lemma}

\begin{IEEEproof}
A brief proof of Lemma \ref{lemma:nbhd d+1 indep symbols, necessary cond for sec, oneshot frmwk} is given \af.
(\ref{eq:d-1, neighborhood key d/d-1 symbols}) implies that, besides $Z_k$, the $|\Nc_k|=d$ \nbrs of \user{k} should collectively hold at least  $d-1$ \indep key symbols. 
This follows from the fundamental result of  \secagg in the server-based centralized topology by Zhao and Sun\cite{zhao2023secure}. Specifically, by viewing \user{k} as a virtual server, securely aggregating the  inputs of its \nbrs   requires $|\Nc_k|-1$  \indep key symbols shared among its \nbrs $\Nc_k$.
Moreover, securely aggregating $W_k$ at any User $i\in \Nc_k$  requires \user{k} to  possess at least $H(W_k)= 1$ key symbol, implying $H(Z_k)\ge 1$. 
Together with (\ref{eq:d-1, neighborhood key d/d-1 symbols}), (\ref{eq:d, neighborhood key d/d-1 symbols})  follows immediately by the chain  rule of entropy, \ie, $H( Z_{\Nc_k  \cup  \{k\} } )=H( Z_{\Nc_k }  |Z_k)  +H(Z_k) \ge (d-1)+1=d$, suggesting each user and its \nbrs should collectively at least $d$ \indep key symbols.
\end{IEEEproof}

Given the linear key design (\ref{eq:indiv key design, one-shot framework}), (\ref{eq:d, neighborhood key d/d-1 symbols}) specializes to $H( \{\Hm_{i,:}N_{1:d}^T\}_{i\in \Nc_k \cup \{k\}} )\ge d$, which implies
\begin{align}
\label{eq:security, rank(H_Nkcupk)=d}
\rank\lf( \Hm_{ \Nc_k \cup  \{k\},:  } \rt)= d, \; \forall  k  \in[K]
\end{align}
(\ref{eq:security, rank(H_Nkcupk)=d}) further implies  $\rank(\Hm) \ge d$, and hence 
$\Amalpha$ must admit a kernel dimension of at least $d$.
Since the \ntlztn vector  $\alphabm$ can be tuned, a necessary condition that jointly captures the   recovery  and  \secty  requirements for the  \achvblty  of any TSA \schm with rates $\rx=1, \rz=1$ and $\rzsigma=d$ is
\be
\label{eq:DMAM kernel dim>=d}
\max\nolimits_{\alphabm \in  \Fmbb_q^{1\times K}   } \dim \ker(\Amalpha) \ge d.
\ee

\tbf{Choice of $\Hm$.}
Suppose there  exists a \modltn vector $ \alphabmstar=(\alpha_1^*, \dots, \alpha_K^* )$ that achieves a kernel \dimtn $  \eta^* \eqdef \dim  \ker(\Amalphastar) \ge d$. Let $\bv_1, \cdots, \bv_{\eta^*} $ be the  $\eta^*$ linearly \indep basis vectors of  $\ker(\Amalphastar)$.  
The key encoding matrix is then chosen  as the first $d$ basis vectors 
\be
\label{eq:H=[b1,...,bd], oneshot frmwk}
\Hm= \lf [\bv_1, \cdots, \bv_d \rt].
\ee 

Given any optimal
\modltn vector $ \alphabmstar=(\alpha_1^*, \dots, \alpha_K^* )$, the  \secty condition in (\ref{eq:neighborhood key d/d-1 symbols}) can be specialized to $\forall k\in [K]$:
\vspace{-.3cm}
\besub
\label{eq:rank(H_Nk) specification to d/d-1}
\begin{align}
& \mathrm{[Security]} \quad  \trm{((\ref{eq:d, neighborhood key d/d-1 symbols})}\Rightarrow)\;  \rank\lf( \Hm_{ \Nc_k \cup \{k\}, :}\rt)   =d,
\label{eq:eq1}\\
& \quad \;
\trm{((\ref{eq:d-1, neighborhood key d/d-1 symbols})}\Rightarrow) \;
\rank\lf( \Hm_{ \Nc_k,:}\rt)  =  
\lf \{  
\begin{array}{ll}
d-1 ,  & \trm{if } \alpha_k^*=0  \\
d,     & \trm{if } \alpha_k^*\ne 0  
\end{array}
\rt.
\label{eq:eq2}
\end{align}
\eesub
where (\ref{eq:eq2})  follows from the observation that, if $ \alpha_k^*=0$,  the  recovery condition (\ref{eq:key neutralization, one-shot framework}) becomes 
\begin{align}
&  \sum\nolimits_{i\in \Nc_k}Z_i  =\lf ( \sum\nolimits_{i\in \Nc_k}\Hm_{i,:}\rt )  N_{1:d}^T=0 \notag\\
&     \Rightarrow \sum\nolimits_{i\in \Nc_k}\Hm_{i,:}=\mbf{0}_{1\times d}    \Rightarrow \rank\lf( \Hm_{\Nc_k,:} \rt)  \le  d-1.
\end{align}
\Bcuz $\rank( H_{ \Nc_k  \cup \{k\}}   )=d$ and $\Hm_{k,:}$ adds at most one more \indep \dimtn to  $ \Hm_{ \Nc_k,:}$, we have  $\rank( \Hm_{\Nc_k,:})  = d-1$. \Inadd, if $\alpha_k^*\ne 0$, (\ref{eq:key neutralization, one-shot framework}) becomes
\begin{align}
\label{eq:recovery relation,alpha_k^* not 0}
 & \alpha_k^*Z_k +     \sum\nolimits_{i\in \Nc_k}Z_i  =\lf (  \alpha_k^* \Hm_{k,: } + \sum\nolimits_{i\in \Nc_k}\Hm_{i,:}\rt )N_{1:d}^T=0 \notag\\
& \hspace{2.5cm}  \Rightarrow \alpha_k^* \Hm_{k,: } + \sum\nolimits_{i\in \Nc_k}\Hm_{i,:}=\mbf{0}_{1\times d}.
\end{align} 
In this case, $\Hm_{k,:}$ lies in the (row) span of  $\Hm_{\Nc_k,:  }$ so that $ \rank( \Hm_{\Nc_k,:})=\rank( \Hm_{\Nc_k \cup  \{k\}  ,:})=d$ according to (\ref{eq:security, rank(H_Nkcupk)=d}).
\Fex,  Example \ref{example:K=6 prism graph}  uses  $\alphabmstar=(2,2,2,2,2,2)\in \Fmbb_5^{1\times 5}$ and
\be 
{\setlength{\arraycolsep}{3.5pt}
\renewcommand{\arraystretch}{0.8}
\Hm=
\begin{bmatrix}
 1 &  0 & 0 & -2 &-1 & -1 \\ 
 0 &  1 & 0 & -1 &-2 & -1 \\ 
 0 &  0 & 1 & -1 &-1 & -2 \\ 
\end{bmatrix}^T \in \Fmbb_5^{6\times  3}.
}
\ee 
Since the optimal \ntlztn \coeffts are all nonzero, it can be verified that $\rank(\Hm_{\Nc_k,:})=3,\forall k\in[6]$. 

To summarize,  under the proposed linear design \frmwk, the recovery (\ie, key \ntlztn) and \secty \reqs (\ref{eq:recovery}) and (\ref{eq:security}) are \resp captured by the kernel condition (\ref{eq: Aalpha*H=0}) on $\Amalpha$ and the rank condition (\ref{eq:rank(H_Nk) specification to d/d-1}) on $\Hm$.
TSA is feasible if  and only if both \reqs are satisfied, \ie,
there exists a \modltn vector $\alphabmstar$ such that the DMAM $\Amalphastar$ admits a kernel of \dimtn at least $d$. \Aar, $\Hm$ is constructed from the kernel of $\Amalphastar$ (\ref{eq:H=[b1,...,bd], oneshot frmwk}). 
\Itf, we present explicit constructions of $\Hm$ for the considered \topos.

\subsection{Specification of $\Hm$ for \Diff Topologies}
\label{subsec:H specification}

\subsubsection{Ring \topo}
Choose a field size $q$  \suth $K\mid (q-1)$. Then $\Fmbb_q$ contains a  primitive \kth root of unity $\omega $, \ie,  $\omega^K=1$, $\omega^m \ne 1, \forall m \in [1:K-1]$. 
Let $ \alphabm=(\alpha, \cdots,  \alpha)$ where $\alpha= -(\omega + \omega^{-1})$. $\Hm$ is chosen as
\be
\label{eq:H,ring, specificiation}
\setlength{\arraycolsep}{3pt}
\renewcommand{\arraystretch}{0.8}
\Hm=
\begin{bmatrix}
1 & \omega & \cdots & \omega^{K-1}\\
1 & \omega^{-1} & \cdots & \omega^{-(K-1)}
\end{bmatrix}^T.
\ee

\subsubsection{Prism graph}
Let $K=2M, M\ge 3$ where $M  \mid (q-1)$.  
Choose a prime number $p \equiv  1\pmod M$.  Let $\omega$ be a primitive $M$-th root of unity and
define $\lambda_t=\omega^{t} + \omega^{-t},
\vv_t=[1,  \omega^t, \cdots, \omega^{(M-1)t}],
t\in \{0,1,M-1\}$ and $\Delta = \lambda_1(\lambda_1-4) \in \Fmbbp $.
If  $\Delta$ is a square in $\Fmbbp$, we work over $\Fmbb_q=\Fmbb_p$; otherwise, 
work over the extension field $ \Fmbb_q= \Fmbb_{p^2}=\Fmbbp[x]/(x^2-\Delta)$. 
Then $\Fmbbq$ contains both a primitive $M$-th root of unity  and a square root of $\Delta$.
Let  $\alphabm =\lf( \alpha_1, \cdots,\alpha_1, \alpha_2, \cdots, \alpha_2      \rt) \in \Fmbb_q^{1\times K}   $ where $\alpha_1, \alpha_2= \frac{-(\lambda+2) \pm  \sqrt{\Delta}  } {2}$. 
$\Hm$ is then chosen as
\be
\label{eq:H,prism graph, specification}
\setlength{\arraycolsep}{3pt}
\renewcommand{\arraystretch}{0.9}
\Hm=
\begin{bmatrix}
\vv_0 & -(\alpha_1 + \lambda_0)\vv_0\\
\vv_1 & -(\alpha_1 + \lambda_1)\vv_1\\
\vv_{M-1} & -(\alpha_1 + \lambda_{M-1})\vv_{M-1}\\
\end{bmatrix}^T.
\ee

\subsubsection{\Cplt graph}
Consider the binary field $\Fmbb_2$. 
Let $\alphabm=(1, \cdots, 1)\in  \Fmbb_2^{1\times K}$. $\Hm$ is chosen as
\be
\label{eq:H,complete graph, specification}
\setlength{\arraycolsep}{3pt}
\renewcommand{\arraystretch}{0.8}
\Hm=\begin{bmatrix}
\mbf{I}_{K-1}\\
\underline{-1}_{1\times (K-1)}
\end{bmatrix},
\ee
where $\mbf{I}_{K-1}$ denotes the $(K-1)\times (K-1)$ identity matrix.

\begin{lemma}
\label{lemma:existence of w, specification}
\tit{For each of the \kgm  given in (\ref{eq:H,ring, specificiation}), (\ref{eq:H,prism graph, specification}) and (\ref{eq:H,complete graph, specification}), there exists a proper field $\Fmbb_q$ and $\omega$ such that $\Hm$ satisfies the rank conditions (\ref{eq:rank(H_Nk) specification to d/d-1}) imposed by  \secty.
}
\end{lemma}

Due to space limitations, the proof of Lemma \ref{lemma:existence of w, specification} is omitted. In general, such proofs rely on probabilistic existence arguments based on the Schwartz-Zippel lemma~\cite{Schwartz}.

\section{Converse}
\label{sec:converse}
We establish lower bounds $\rx\ge 1, \rz\ge 1$ and $\rzsigma\ge d$ on the \comm  and secret key rates for any $d$-\regu network. Since these bounds match the achievable rates derived in Section \ref{sec:proposed scheme}, the optimal rate region is characterized.

\subsubsection{Proof of $\rx \ge 1 $ and $\rz\ge 1$}
The converse proof for the \comm and \indiv key rates are relatively standard:
(i) A cut-set bound for recovering $\sum_{i\in \Nc_k}W_i$ at any User $k$ requires User $i \in \Nc_k$ to transmit at least $H(W_i)=L$ symbols; otherwise $ W_i$ cannot propagate through the network. This implies $\rx\ge 1$. (ii) For any pair of neighboring users $(i,j)$, protecting $W_i$ (embedded in $X_i$) from User $j$ requires $Z_i$ to contain at least $H(W_i)=L$ \iid uniform symbols, yielding $\rz\ge 1$, according to Shannon's one-time pad (OTP) encryption theorem~\cite{shannon1949communication}.

\subsubsection{Proof of $\rzsigma \ge d$}
Consider any \user{k} with  \nbr set $\Nc_k$. 
Lemma \ref{lemma:nbhd d+1 indep symbols, necessary cond for sec, oneshot frmwk} (see Section \ref{sec:linear design frmwk}) suggests that $H(Z_{\Nc_k}|Z_k) \ge d-1,\forall k\in[K]$. Moreover, it is easy to see $H(Z_k)\ge L$ as proved in the previous  subsection.
\Thf, 
\begin{align}
 H(\zsigma) &  \overset{(\ref{eq:key gen from source key})}{\ge } H(Z_{\Nc_k}, Z_k)=H(Z_k) + H(Z_{\Nc_k}|Z_k) \overset{\trm{(\ref{eq:d-1, neighborhood key d/d-1 symbols})}}{\ge } d,\notag\\
& \Rightarrow  \rzsigma \eqdef {\lzsigma}/{L} \ge  {H(\zsigma)}/{L} \ge d. \;\; (L=1)
\end{align}

\section{Conclusion}
\label{sec:conclusion}
In this work, we introduced \tsa and proposed a unified linear design framework that achieves the extremal rate region $\Rc^*_{\rm ext.}=\{(\rx,\rz ,\rzsigma): \rx\ge 1, \rz \ge 1, \rzsigma  \ge d  \}$ for $d$-regular networks by linking TSA achievability to the kernel dimension of the diagonally modulated adjacency matrix (DMAM) of the underlying communication graph. We showed that TSA is feasible whenever the DMAM admits at least $d$  independent kernel \dimtns. We further identified several  classes of $d$-regular networks that \achv $\Rc^*_{\rm ext.}$, together with explicit secret key designs. It  should be noted that many additional $d$-\regu \ntwks can \achv $\Rc^*_{\rm ext.}$ but are not enumerated  due to space limitations.
Future work includes characterizing which $d$-regular networks can or cannot achieve $\Rc^*_{\rm ext.}$, as well as determining the optimal rate region for arbitrary network topologies using  graph spectral and algebraic tools.

\clearpage

\bibliographystyle{IEEEtran}
\bibliography{references_secagg.bib}

\end{document}